%% file: arxiv-wrapper.tex
\documentclass[a4paper, 11pt]{article}
\usepackage[utf8]{inputenc}
\usepackage[T1]{fontenc}
\usepackage{amsmath,amssymb,array}
\usepackage{booktabs}

\usepackage{graphicx}

\providecommand{\tightlist}{%
  \setlength{\itemsep}{0pt}\setlength{\parskip}{0pt}}

\RequirePackage{geometry}
\RequirePackage{fancyhdr}
\RequirePackage{fancyvrb}
\RequirePackage{alltt}

\RequirePackage[hyphens]{url}
\RequirePackage[pagebackref]{hyperref}

\let\pkg=\strong
\newcommand{\CRANpkg}[1]{\href{http://CRAN.R-project.org/package=#1}{\pkg{#1}}}%

\usepackage{natbib}

\newcommand{\address}[1]{\addvspace{\baselineskip}\noindent\emph{#1}}

\newenvironment{CSLReferences}%
  {}%
  {\par}
  
\makeatletter
\renewcommand{\@seccntformat}[1]{}
\makeatother

\RequirePackage{xcolor}
\definecolor{link}{rgb}{0.45,0.51,0.67}
\hypersetup{
  colorlinks,%
  citecolor=link,%
  filecolor=link,%
  linkcolor=link,%
  urlcolor=link
}

\DefineVerbatimEnvironment{Sinput}{Verbatim}{fontsize=\small}
\DefineVerbatimEnvironment{Soutput}{Verbatim}{fontsize=\small}
\DefineVerbatimEnvironment{Scode}{Verbatim}{fontsize=\small}
\DefineVerbatimEnvironment{Sin}{Verbatim}{fontsize=\small}
\DefineVerbatimEnvironment{Sout}{Verbatim}{fontsize=\small}

\begin{document}

\input{paper-r01}

\end{document}

%% file: paper-r01.tex
% !TeX root = RJwrapper.tex
\title{Computer Algebra in R Bridges a Gap Between Mathematics and Data in the Teaching of Statistics and Data Science}
\author{Mikkel Meyer Andersen and Søren Højsgaard}

\maketitle

\address{%
Mikkel Meyer Andersen\\
Department of Mathematical Sciences, Aalborg University, Denmark\\%
Skjernvej 4A\\ 9220 Aalborg Ø, Denmark\\
\textit{ORCiD: \href{https://orcid.org/0000-0002-0234-0266}{0000-0002-0234-0266}}\\%
\href{mailto:mikl@math.aau.dk}{\nolinkurl{mikl@math.aau.dk}}%
}

\address{%
Søren Højsgaard\\
Department of Mathematical Sciences, Aalborg University, Denmark\\%
Skjernvej 4A\\ 9220 Aalborg Ø, Denmark\\
\textit{ORCiD: \href{https://orcid.org/0000-0002-3269-9552}{0000-0002-3269-9552}}\\%
\href{mailto:sorenh@math.aau.dk}{\nolinkurl{sorenh@math.aau.dk}}%
}

\abstract{%
The capability of R to do symbolic mathematics is enhanced by the \texttt{caracas} package. This package uses the Python computer algebra library SymPy as a back-end but \texttt{caracas} is tightly integrated in the R environment. This enables the R user with symbolic mathematics within R at a high abstraction level rather than using text strings and text string manipulation as the case would be if using SymPy from R directly. We demonstrate how mathematics and statistics can benefit from bridging computer algebra and data via R. This is done thought a number of examples and we propose some topics for small student projects. The \texttt{caracas} package integrates well with e.g.~\texttt{Rmarkdown}, and as such creation of scientific reports and teaching is supported.
}

\hypertarget{introduction}{%
\section{Introduction}\label{introduction}}

The \CRANpkg{caracas} package (Andersen and Højsgaard 2021) and the \CRANpkg{Ryacas}
package (Andersen and Højsgaard 2019) enhance the capability of R to handle
symbolic mathematics. In this paper we will illustrate the use of the
\texttt{caracas} package (version 2.0.1) in connection with teaching mathematics and
statistics. Focus is on 1) treating statistical models symbolically,
2) on bridging the gap between symbolic mathematics and numerical
computations and 3) on preparing teaching material in a reproducible
framework (provided by, e.g.~\CRANpkg{rmarkdown} (Allaire et al. 2021; Xie, Allaire, and Grolemund 2018; Xie, Dervieux, and Riederer 2020)). The \texttt{caracas} package
is available from CRAN. The open-source development version of
\texttt{caracas} is available at \url{https://github.com/r-cas/caracas} and
readers are recommended to study the online documentation at
\url{https://r-cas.github.io/caracas/}. The \texttt{caracas} package provides an
interface from R to the Python package SymPy (Meurer et al. 2017). This
means that SymPy is ``running under the hood'' of R via the
\texttt{reticulate} package (Ushey, Allaire, and Tang 2020). The SymPy package is mature and
robust with many users and developers.

The benefit of using \texttt{caracas} instead of using SymPy via \texttt{reticulate}
is that work is performed at a higher abstraction level in a session of
operations and not coding at a lower-level using text strings and
text string manipulation.

Neither \texttt{caracas} nor \texttt{Ryacas} are as powerful as some
of the larger commercial computer algebra systems (CAS). The virtue of
\texttt{caracas} and \texttt{Ryacas} lie elsewhere:
(1) Mathematical tools like equation solving, summation, limits, symbolic linear
algebra, outputting in tex format etc. are directly available from
within R.
(2) The packages enable working with the same language and in the same
environment as the user does for statistical analyses.
(3) Symbolic mathematics can easily be combined with data which is
helpful in e.g.~numerical optimization.
(4) The packages are open-source and therefore support e.g.~education - also for people
with limited economical means and thus contributing to United
Nations sustainable development goals (United Nations General Assembly 2015).

The paper is organized in the following sections: The section
\protect\hyperlink{introducing-caracas}{Introducing \texttt{caracas}} briefly introduces
the \texttt{caracas} package and its syntax, including how \texttt{caracas} can be
used in connection with preparing texts, e.g.~teaching material. More
details are provided in \protect\hyperlink{appendix}{Appendix}.
Several vignettes illustrating \texttt{caracas} are provided and they are
also available online, see \url{https://r-cas.github.io/caracas/}. The
section \protect\hyperlink{statistics-examples}{Statistics examples} is the main section of the paper and
here we present a sample of statistical models where we believe that a
symbolic treatment is a valuable supplement to a numerical in
connection with teaching. The section {[}Possible topics to study{]}
contains suggestions about hand-on activities for students. Lastly,
the section \protect\hyperlink{discussion-and-future-work}{Discussion and future work} contains a discussion of the
paper.

\hypertarget{introducing-caracas}{%
\section{\texorpdfstring{Introducing \texttt{caracas}}{Introducing caracas}}\label{introducing-caracas}}

Introduce key concepts and show functionality subsequently needed in the section \protect\hyperlink{statistics-examples}{Statistics examples}.

\hypertarget{documents-with-mathematical-content}{%
\subsection{Documents with mathematical content}\label{documents-with-mathematical-content}}

A LaTeX rendering of a \texttt{caracas} symbol, say \texttt{x} is obtained by typing
\texttt{\$\$x\ =\ \textasciigrave{}r\ tex(x)\textasciigrave{}\$\$}. This feature is useful
when creating documents with a mathematical content and has been used
extensively throughout this paper (looks nice and saves space).

\hypertarget{symbols}{%
\subsection{Symbols}\label{symbols}}

A \texttt{caracas} symbol is a list with a \texttt{pyobj} slot and the class
\texttt{caracas\_symbol}. The \texttt{pyobj} is a Python object (often a SymPy
object). As such, a \texttt{caracas} symbol (in R) provides a handle to a
Python object. In the design of \texttt{caracas} we have tried to make
this distinction something the user should not be concerned with, but
it is worthwhile being aware of the distinction. Whenever we refer to
a symbol we mean a \texttt{caracas} symbol. Two functions that create
symbols are \texttt{def\_sym()} and \texttt{as\_sym()}; these and other functions that
create symbols will be illustrated below.

\hypertarget{linear-algebra}{%
\subsection{Linear algebra}\label{linear-algebra}}

We create a symbolic matrix from an R object and a symbolic
vector directly. A vector is a one-column matrix which is printed as
its transpose to save space. Matrix products are computed using the
\texttt{\%*\%} operator:

\begin{verbatim}
R> M0 <- toeplitz(c("a", "b"))  ## Character matrix
R> M  <- as_sym(M0)             ## as_sym() converts to a caracas symbol
R> v  <- vector_sym(2, "v")     ## vector_sym creates symbolic vector
R> y  <- M %*% v
R> Minv <- inv(M) %>% simplify()
R> v2 <- Minv %*% y  |> simplify()
\end{verbatim}

Default printing of \texttt{M} is

\begin{verbatim}
R> M
\end{verbatim}

\begin{verbatim}
#> [c]: [a  b]
#>      [    ]
#>      [b  a]
\end{verbatim}

while the LaTeX rendering of the symbols above are:

\[
M = \left[\begin{matrix}a & b\\b & a\end{matrix}\right]; \; 
v = \left[\begin{matrix}v_{1}\\v_{2}\end{matrix}\right]; \;
y = \left[\begin{matrix}a v_{1} + b v_{2}\\a v_{2} + b v_{1}\end{matrix}\right]; \;
M^{-1} = \left[\begin{matrix}\frac{a}{a^{2} - b^{2}} & - \frac{b}{a^{2} - b^{2}}\\- \frac{b}{a^{2} - b^{2}} & \frac{a}{a^{2} - b^{2}}\end{matrix}\right]; \;
v2 = \left[\begin{matrix}v_{1}\\v_{2}\end{matrix}\right] . 
\]

The determinant of \(M\), \(det(M)=a^2 - b^2\), can be factored out of
the matrix by dividing each entry with the determinant and multiplying
the new matrix by the determinant which simplifies the appearance of
the matrix:

\begin{verbatim}
R> Minv_fact <- as_factor_list(1 / det(M), simplify(det(M) * Minv))
\end{verbatim}

Hence we have in LaTeX format:

\[
\quad M^{-1} = \frac{1}{a^{2} - b^{2}}  \left[\begin{matrix}a & - b\\- b & a\end{matrix}\right] = \left[\begin{matrix}\frac{a}{a^{2} - b^{2}} & - \frac{b}{a^{2} - b^{2}}\\- \frac{b}{a^{2} - b^{2}} & \frac{a}{a^{2} - b^{2}}\end{matrix}\right] .
\]

A \texttt{caracas} symbol can be coerced to an R expression
using \texttt{as\_expr()}.
Symbols can be substituted with other symbols or with numerical values
using \texttt{subs()}:

\begin{verbatim}
R> as_expr(M)
\end{verbatim}

\begin{verbatim}
#> expression(matrix(c(a, b, b, a), nrow = 2))
\end{verbatim}

\begin{verbatim}
R> def_sym(a) ## This creates the symbol 'a'
R> a
\end{verbatim}

\begin{verbatim}
#> [c]: a
\end{verbatim}

\begin{verbatim}
R> M2 <- subs(M, "b", "a^2")
R> M3 <- subs(M2, a, 2)
\end{verbatim}

\[
M2 = \left[\begin{matrix}a & a^{2}\\a^{2} & a\end{matrix}\right]; \quad
M3 = \left[\begin{matrix}2 & 4\\4 & 2\end{matrix}\right].
\]

\hypertarget{calculus}{%
\subsection{Calculus}\label{calculus}}

Next, we define a \texttt{caracas} symbol \texttt{x} and
subsequently a \texttt{caracas} polynomial \texttt{p} in \texttt{x} (\texttt{p} becomes a symbol because \texttt{x} is):

\begin{verbatim}
R> def_sym(x)  
R> p <- 1 - x^2 + x^3 + x^4/4 - 3 * x^5 / 5 + x^6 / 6
\end{verbatim}

We investigate \texttt{p} further by finding the gradient and Hessian of \texttt{p}. The gradient factors which shows that the stationary
points are \(-1\), \(0\), \(1\) and \(2\):

\begin{verbatim}
R> g <- der(p, x) 
R> g2 <- factor_(g)
R> h <- der2(p, x)
\end{verbatim}

Notice here: Several functions have a postfix underscore as a simple
way of distinguishing them from R functions with a different
meaning.

\[
 \texttt{g}  = x^{5} - 3 x^{4} + x^{3} + 3 x^{2} - 2 x; \quad 
 \texttt{g2}  = x \left(x - 2\right) \left(x - 1\right)^{2} \left(x + 1\right).
\]

In a more general setting we can find the stationary points by equating the gradient to zero:
The output \texttt{sol} is a list of solutions in which each solution is a list of \texttt{caracas} symbols.

\begin{verbatim}
R> sol <- solve_sys(lhs = g, rhs = 0, vars = x)
R> sol
\end{verbatim}

\begin{verbatim}
#> Solution 1:
#>   x =  -1 
#> Solution 2:
#>   x =  0 
#> Solution 3:
#>   x =  1 
#> Solution 4:
#>   x =  2
\end{verbatim}

\begin{verbatim}
R> sol_expr <- sapply(sol, sapply, as_expr) |> unname()
R> sol_expr
\end{verbatim}

\begin{verbatim}
#> [1] -1  0  1  2
\end{verbatim}

\begin{figure}
\centering
\includegraphics{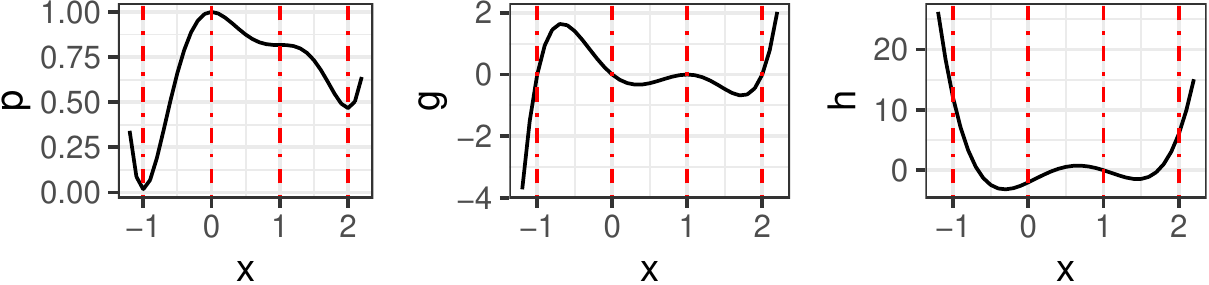}
\caption{\label{fig:calculus}Left: A polynomium. Center: The gradient. Right: The Hessian.}
\end{figure}

A \texttt{caracas} symbol can be turned into an R function for subsequent
numerical evaluation using \texttt{as\_func()}, see
Fig. \ref{fig:calculus}.
The stationary points are indicated in the plots.

\begin{verbatim}
R> p_fn <- as_func(p)
R> p_fn
\end{verbatim}

\begin{verbatim}
#> function (x) 
#> {
#>     x^6/6 - 3 * x^5/5 + x^4/4 + x^3 - x^2 + 1
#> }
#> <environment: 0x564438d316a8>
\end{verbatim}

\begin{verbatim}
R> g_fn <- as_func(g)
R> h_fn <- as_func(h)
R> h_fn(sol_expr)
\end{verbatim}

\begin{verbatim}
#> [1] 12 -2  0  6
\end{verbatim}

The sign of the Hessian in the stationary points shows that \(-1\) and
\(2\) are local minima, \(0\) is a local maximum and \(1\) is an inflection
point.

\hypertarget{integration}{%
\subsection{Integration}\label{integration}}

The unit circle is given by \(x^2 + y^2 = 1\) so the area of the upper
half of the unit circle is \(\int_{-1}^1 \sqrt{1-x^2}\; dx\) (which is
known to be \(\pi/2\)). This result is produced by \texttt{caracas} while the
\texttt{integrate} function in R produces the approximate result \(1.57\).

\begin{verbatim}
R> x <- as_sym("x")
R> half_circle_ <- sqrt(1-x^2)
R> ad <- int(half_circle_, "x")          ## Anti derivative
R> area <- int(half_circle_, "x", -1, 1) ## Definite integral
\end{verbatim}

\[
\texttt{ad} = \frac{x \sqrt{1 - x^{2}}}{2} + \frac{\operatorname{asin}{\left(x \right)}}{2}; \quad
\texttt{area} = \frac{\pi}{2}.
\]

\hypertarget{unevaluated-expressions}{%
\subsection{Unevaluated expressions}\label{unevaluated-expressions}}

Finally, we illustrate creation of unevaluated expressions:

\begin{verbatim}
R> def_sym(x, n)
R> y <- (1 + x/n)^n
R> l <- lim(y, n, Inf, doit = FALSE)
R> l_2 <- doit(l)
\end{verbatim}

\[
l = \lim_{n \to \infty} \left(1 + \frac{x}{n}\right)^{n}; \quad l_2 = e^{x}
\]

Several functions have the \texttt{doit} argument, e.g.~\texttt{lim()}, \texttt{int()} and \texttt{sum\_()}.
Unevaluated expressions help making reproducible documents where the changes
in code appears automatically in the generated formulas.

\hypertarget{statistics-examples}{%
\section{Statistics examples}\label{statistics-examples}}

In this section we examine larger statistical examples and
demonstrate how \texttt{caracas} can help improve understanding of the models.

\hypertarget{example-linear-models}{%
\subsection{Example: Linear models}\label{example-linear-models}}

A matrix algebra approach to e.g.~linear models is very clear and
concise. On the other hand, it can also be argued that matrix algebra
obscures what is being computed. Numerical examples are useful for
some aspects of the computations but not for other. In this respect
symbolic computations can be enlightening.

Consider a two-way analysis of variance (ANOVA) with one observation
per group, see Table \ref{tab:anova-two-way-table}.

\begin{table}[!h]

\caption{\label{tab:anova-two-way-table}Two-by-two layout of data.}
\centering
\begin{tabular}[t]{|>{}l|>{}l|}
\hline
$y_{11}$ & $y_{12}$\\
\hline
$y_{21}$ & $y_{22}$\\
\hline
\end{tabular}
\end{table}

\begin{verbatim}
R> nr <- 2
R> nc <- 2
R> y  <- as_sym(c("y_11", "y_21", "y_12", "y_22"))
R> dat <- expand.grid(r = factor(1:nr), s = factor(1:nc))
R> X <- model.matrix(~ r + s, data = dat) |> as_sym()
R> b <- vector_sym(ncol(X), "b")
R> mu <- X %*% b
\end{verbatim}

For the specific model we have random variables \(y=(y_{ij})\). All
\(y_{ij}\)s are assumed independent and \(y_{ij}\sim N(\mu_{ij}, v)\).
The corresponding mean vector \(\mu\) has the form given below:

\[
y = \left[\begin{matrix}y_{11}\\y_{21}\\y_{12}\\y_{22}\end{matrix}\right], \quad X=\left[\begin{matrix}1 & . & .\\1 & 1 & .\\1 & . & 1\\1 & 1 & 1\end{matrix}\right], \quad b=\left[\begin{matrix}b_{1}\\b_{2}\\b_{3}\end{matrix}\right], \quad  \mu = X b = \left[\begin{matrix}b_{1}\\b_{1} + b_{2}\\b_{1} + b_{3}\\b_{1} + b_{2} + b_{3}\end{matrix}\right] .
\]

Above and elsewhere, dots represent zero. This is obtained with the \texttt{zero\_as\_dot} argument to the \texttt{tex()} function.
The least squares estimate of \(b\) is the vector \(\hat{b}\) that minimizes \(||y-X b||^2\) which leads to the normal equations \((X^\top X)b = X^\top y\) to be solved. If \(X\) has full rank, the unique solution to the normal
equations is \(\hat{b} = (X^\top X)^{-1} X^\top y\). Hence the
estimated mean vector is \(\hat \mu = X\hat{b}=X(X^\top X)^{-1} X^\top y\). Symbolic computations are
not needed for quantities involving only the model matrix \(X\), but
when it comes to computations involving \(y\), a symbolic treatment of
\(y\) is useful:

\begin{verbatim}
R> XtX <- t(X) %*% X
R> XtXinv <- inv(XtX)
R> Xty <- t(X) %*% y
R> b_hat <- XtXinv %*% Xty
\end{verbatim}

\begin{align}
X^\top y &= \left[\begin{matrix}y_{11} + y_{12} + y_{21} + y_{22}\\y_{21} + y_{22}\\y_{12} + y_{22}\end{matrix}\right]; \quad 
\quad
\hat{b} = \frac{1}{2}  \left[\begin{matrix}\frac{3 y_{11}}{2} + \frac{y_{12}}{2} + \frac{y_{21}}{2} - \frac{y_{22}}{2}\\- y_{11} - y_{12} + y_{21} + y_{22}\\- y_{11} + y_{12} - y_{21} + y_{22}\end{matrix}\right].
\end{align}

Hence \(X^\top y\) (a sufficient reduction of data if the variance is
known) consists of the sum of all observations, the sum of
observations in the second row and the sum of observations in the
second column. For \(\hat{b}\), the second component is, apart from a
scaling, the sum of the second row minus the sum of the first
row. Likewise, the third component is the sum of the second column
minus the sum of the first column. Hence, for example the second
component of \(\hat{b}\) is the difference in mean between the first and
second column in Table \ref{tab:anova-two-way-table}.

\hypertarget{example-logistic-regression}{%
\subsection{Example: Logistic regression}\label{example-logistic-regression}}

In the following we go through details of a logistic regression model,
see e.g. McCullagh and Nelder (1989) for a classical description of logistic
regression.

\begin{figure}
\centering
\includegraphics{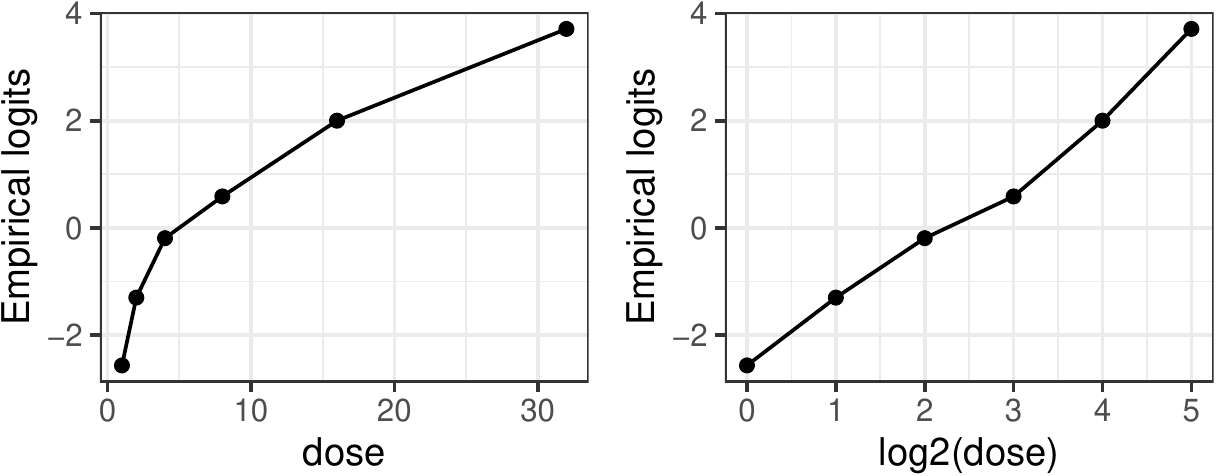}
\caption{\label{fig:budworm}Insecticide mortality of the moth tobacco budworm.}
\end{figure}

As an example, consider the \texttt{budworm} data from the \CRANpkg{doBy} package (Højsgaard and Halekoh 2023).
The data shows the number of killed moth tobacco budworm
\emph{Heliothis virescens}. Batches of 20 moths of each sex were
exposed for three days to the pyrethroid and the number in each batch
that were dead or knocked down was recorded.
Below we focus only on male budworms and the mortality is illustrated
in Figure \ref{fig:budworm} (produced with \CRANpkg{ggplot2} (Wickham 2016)). On the \(y\)-axis we have the empirical
logits, i.e.~\(\log((\text{ndead} + 0.5)/(\text{ntotal}-\text{ndead} + 0.5))\). The figure suggests that logit grows linearly with log dose.

\begin{verbatim}
R> data(budworm, package = "doBy")
R> bud <- subset(budworm, sex == "male")
R> bud
\end{verbatim}

\begin{verbatim}
#>    sex dose ndead ntotal
#> 1 male    1     1     20
#> 2 male    2     4     20
#> 3 male    4     9     20
#> 4 male    8    13     20
#> 5 male   16    18     20
#> 6 male   32    20     20
\end{verbatim}

Observables are binomially distributed, \(y_i \sim \text{bin}(p_i, n_i)\). The probability \(p_i\) is connected to a \(q\)-vector of
covariates \(x_i=(x_{i1}, \dots, x_{iq})\) and a \(q\)-vector of
regression coefficients \(b=(b_1, \dots, b_q)\) as follows: The term
\(s_i = x_i \cdot b\) is denoted the \emph{linear predictor}. The
probability \(p_i\) can be linked to \(s_i\) in different ways, but the
most commonly employed is via the \emph{logit link function} which is
\(\text{logit}(p_i) = \log(p_i/(1-p_i))\) so here \(\text{logit}(p_i) = s_i\). Based on Figure \ref{fig:budworm}, we consider the specific
model with \(s_i = b_1 + b_2 \log2(dose_i)\). For later use, we define the data matrix below:

\begin{verbatim}
R> DM <- cbind(model.matrix(~log2(dose), data=bud),
+             bud[, c("ndead", "ntotal")])  |> as.matrix()
R> DM |> head(3)
\end{verbatim}

\begin{verbatim}
#>   (Intercept) log2(dose) ndead ntotal
#> 1           1          0     1     20
#> 2           1          1     4     20
#> 3           1          2     9     20
\end{verbatim}

\hypertarget{each-component-of-the-likelihood}{%
\subsubsection{Each component of the likelihood}\label{each-component-of-the-likelihood}}

The log-likelihood is \(\log L=\sum_i y_i \log(p_i) + (n_i-y_i) \log(1-p_i) = \sum_i \log L_i\), say.
Consider the contribution to the total log-likelihood from the \(i\)th
observation which is \(\log L_i = l_i = y_i \log(p_i) + (n_i-y_i) \log(1-p_i)\).
Since we are focusing on one observation only, we shall ignore the
subscript \(i\) in this section. First notice that with
\(s = \log(p/(1-p))\) we can find \(p\) as:

\begin{verbatim}
R> def_sym(s, p)
R> sol_ <- solve_sys(lhs = log(p / (1 - p)), rhs = s, vars = p)
R> p_s <- sol_[[1]]$p
\end{verbatim}

\[
\texttt{p\_s} = \frac{e^{s}}{e^{s} + 1}
\]

Next, find the likelihood as a function of \(p\), as a function of \(s\)
and as a function of \(b\). The underscore in \texttt{logLb\_} and elsewhere
indicates that this expression is defined in terms of other
symbols. The log-likelihood can be maximized using e.g.~Newton-Rapson
(see e.g. Nocedal and Wright (2006)) and in this connection we need the score function,
\(S\), and the Hessian, \(H\):

\begin{verbatim}
R> def_sym(y, n)
R> b  <- vector_sym(2, "b")
R> x  <- vector_sym(2, "x")
R> logLp_ <- y * log(p) + (n - y) * log(1 - p) ## logL as fn of p
R> s_b <- sum(x * b)                           ## s as fn of b
R> p_b <- subs(p_s, s, s_b)                    ## p as fn of b
R> logLb_ <- subs(logLp_, p, p_b)              ## logL as fn of b
R> Sb_ <- score(logLb_, b) |> simplify()
R> Hb_ <- hessian(logLb_, b) |> simplify()
\end{verbatim}

\begin{align}
\texttt{p\_b}   &= \frac{e^{b_{1} x_{1} + b_{2} x_{2}}}{e^{b_{1} x_{1} + b_{2} x_{2}} + 1}, \\
\texttt{logLb}\_ &= y \log{\left(\frac{e^{b_{1} x_{1} + b_{2} x_{2}}}{e^{b_{1} x_{1} + b_{2} x_{2}} + 1} \right)} + \left(n - y\right) \log{\left(1 - \frac{e^{b_{1} x_{1} + b_{2} x_{2}}}{e^{b_{1} x_{1} + b_{2} x_{2}} + 1} \right)}, \\
\texttt{Sb}\_    &= \left[\begin{matrix}\frac{x_{1} \left(- n e^{b_{1} x_{1} + b_{2} x_{2}} + y e^{b_{1} x_{1} + b_{2} x_{2}} + y\right)}{e^{b_{1} x_{1} + b_{2} x_{2}} + 1}\\\frac{x_{2} \left(- n e^{b_{1} x_{1} + b_{2} x_{2}} + y e^{b_{1} x_{1} + b_{2} x_{2}} + y\right)}{e^{b_{1} x_{1} + b_{2} x_{2}} + 1}\end{matrix}\right], \\
\texttt{Hb}\_    &= \left[\begin{matrix}- \frac{n x_{1}^{2} e^{b_{1} x_{1} + b_{2} x_{2}}}{2 e^{b_{1} x_{1} + b_{2} x_{2}} + e^{2 b_{1} x_{1} + 2 b_{2} x_{2}} + 1} & - \frac{n x_{1} x_{2} e^{b_{1} x_{1} + b_{2} x_{2}}}{2 e^{b_{1} x_{1} + b_{2} x_{2}} + e^{2 b_{1} x_{1} + 2 b_{2} x_{2}} + 1}\\- \frac{n x_{1} x_{2} e^{b_{1} x_{1} + b_{2} x_{2}}}{2 e^{b_{1} x_{1} + b_{2} x_{2}} + e^{2 b_{1} x_{1} + 2 b_{2} x_{2}} + 1} & - \frac{n x_{2}^{2} e^{b_{1} x_{1} + b_{2} x_{2}}}{2 e^{b_{1} x_{1} + b_{2} x_{2}} + e^{2 b_{1} x_{1} + 2 b_{2} x_{2}} + 1}\end{matrix}\right] . 
\end{align}

There are various possible approaches from here when it comes
maximizing the total log likelihood. One is to insert data case by
case into the symbolic log likelihood. This yields a list of new \texttt{caracas}
symbol which depends on the unknown regression parameters:

\begin{verbatim}
R> nms <- c("x1", "x2", "y", "n")
R> DM_lst <- doBy::split_byrow(DM)
R> logLb_lst <- lapply(DM_lst, function(vls) {
+     subs(logLb_, nms, vls)
+ })
\end{verbatim}

For example, the contribution from the third observation to the total log likelihood is:

\begin{align}
\texttt{logLb\_lst[[3]]}  &= 9 \log{\left(\frac{e^{b_{1} + 2 b_{2}}}{e^{b_{1} + 2 b_{2}} + 1} \right)} + 11 \log{\left(1 - \frac{e^{b_{1} + 2 b_{2}}}{e^{b_{1} + 2 b_{2}} + 1} \right)}.
\end{align}

These
symbols can be added up and the sum can be maximized either e.g.\\
using SymPy (not pursued here) or by converting the sum to an R
function which can be maximized using one of R's internal
optimization procedures:

\begin{verbatim}
R> logLb_tot <- Reduce(`+`, logLb_lst) 
R> logLb_fn  <- as_func(logLb_tot, vec_arg = TRUE)
R> opt <- optim(c(b1=0, b2=0), logLb_fn, control = list(fnscale = -1), hessian = TRUE)
R> opt$par
\end{verbatim}

\begin{verbatim}
#>    b1    b2 
#> -2.82  1.26
\end{verbatim}

The same model can be fitted e.g.~using R's \texttt{glm()} function as follows (output omitted):

\begin{verbatim}
R> m <- glm(cbind(ndead, ntotal - ndead) ~ log2(dose), family=binomial(), data=bud)
R> m |> coef()
\end{verbatim}

\begin{verbatim}
#> (Intercept)  log2(dose) 
#>       -2.82        1.26
\end{verbatim}

\hypertarget{the-total-likelihood-symbolically}{%
\subsubsection{The total likelihood symbolically}\label{the-total-likelihood-symbolically}}

We conclude this section by illustrating that the log-likelihood for the entire dataset
can be constructed in a few steps (output is omitted to save space):

\begin{verbatim}
R> N <- 6; q <- 2
R> X <- matrix_sym(N, q, "x")
R> n <- vector_sym(N, "n")
R> y <- vector_sym(N, "y")
R> p <- vector_sym(N, "p")
R> s <- vector_sym(N, "s")
R> b <- vector_sym(q, "b")
\end{verbatim}

\[
 X=\left[\begin{matrix}x_{11} & x_{12}\\x_{21} & x_{22}\\x_{31} & x_{32}\\x_{41} & x_{42}\\x_{51} & x_{52}\\x_{61} & x_{62}\end{matrix}\right], \quad
 n=\left[\begin{matrix}n_{1}\\n_{2}\\n_{3}\\n_{4}\\n_{5}\\n_{6}\end{matrix}\right], \quad
 y=\left[\begin{matrix}y_{1}\\y_{2}\\y_{3}\\y_{4}\\y_{5}\\y_{6}\end{matrix}\right] .
\]

The symbolic computations are as follows: We express the linear predictor \(s\) as function of the regression coefficients \(b\) and express the probability \(p\) as function of the linear predictor:

\begin{verbatim}
R> logLp <- sum(y * log(p) + (n - y) * log(1 - p)) ## logL as fn of p
R> p_s <- exp(s) / (exp(s) + 1)                    ## p as fn of s
R> s_b <- X %*% b                                  ## s as fn of b
R> p_b <- subs(p_s, s, s_b)                        ## p as fn of b
R> logLb_ <- subs(logLp, p, p_b)                   ## logL as fn of b
\end{verbatim}

Next step could be to go from symbolic to numerical computations by
inserting numerical values. From here, one may proceed by computing
the score function and the Hessian matrix and solve the score
equation, using e.g.~Newton-Rapson. Alternatively, one might create an
R function based on the log-likelihood, and maximize this function
using one of R's optimization methods (see the example in the
previous section):

\begin{verbatim}
R> logLb <- subs(logLb_, cbind(X, y, n), DM)
R> logLb_fn <- as_func(logLb, vec_arg = TRUE)
R> opt <- optim(c(b1=0, b2=0), logLb_fn, control = list(fnscale = -1), hessian = TRUE)
R> opt$par
\end{verbatim}

\begin{verbatim}
#>    b1    b2 
#> -2.82  1.26
\end{verbatim}

\hypertarget{example-constrained-maximum-likelihood}{%
\subsection{Example: Constrained maximum likelihood}\label{example-constrained-maximum-likelihood}}

In this section we illustrate constrained optimization using Lagrange multipliers.
This is demonstrated for the independence model for a two-way contingency table.
Consider a \(2 \times 2\) contingency table with cell
counts \(y_{ij}\) and cell probabilities \(p_{ij}\) for \(i=1,2\) and \(j=1,2\),
where \(i\) refers to row and \(j\) to column as
illustrated in Table \ref{tab:anova-two-way-table}.

Under multinomial sampling, the log likelihood is
\[
 l = \log L = \sum_{ij} y_{ij} \log(p_{ij}).
\]

Under the assumption of independence between rows and columns, the cell
probabilities have the form, (see e.g. Højsgaard, Edwards, and Lauritzen (2012), p.~32)
\[
p_{ij}=u \cdot r_i \cdot s_j.
\]

To make the parameters \((u, r_i, s_j)\) identifiable, constraints
must be imposed. One possibility is to require that \(r_1=s_1=1\). The
task is then to estimate \(u\), \(r_2\), \(s_2\) by maximizing the log likelihood
under the constraint that \(\sum_{ij} p_{ij} = 1\). These constraints
can be
imposed using a Lagrange multiplier where we solve the
unconstrained optimization problem \(\max_p Lag(p)\) where
\begin{align}
  Lag(p) &= -l(p) + \lambda g(p) \quad \text{under the constraint that} \\
  g(p) &= \sum_{ij} p_{ij} - 1 = 0 ,
\end{align}
where \(\lambda\) is a Lagrange multiplier.
In SymPy, \texttt{lambda} is a reserved symbol. Hence the underscore as postfix below:

\begin{verbatim}
R> def_sym(u, r2, s2, lambda_)
R> y  <- as_sym(c("y_11", "y_21", "y_12", "y_22"))
R> p  <- as_sym(c("u", "u*r2", "u*s2", "u*r2*s2"))
R> logL <- sum(y * log(p))
R> Lag  <- -logL + lambda_ * (sum(p) - 1) 
R> vars <- list(u, r2, s2, lambda_)
R> gLag <- der(Lag, vars)
R> sol  <- solve_sys(gLag, vars)
R> print(sol, method = "ascii")
\end{verbatim}

\begin{verbatim}
#> Solution 1:
#>   lambda_ =  y_11 + y_12 + y_21 + y_22 
#>   r2      =  (y_21 + y_22)/(y_11 + y_12) 
#>   s2      =  (y_12 + y_22)/(y_11 + y_21) 
#>   u       =  (y_11 + y_12)*(y_11 + y_21)/(y_11 + y_12 + y_21 + y_22)^2
\end{verbatim}

\begin{verbatim}
R> sol <- sol[[1]]
\end{verbatim}

There is only one critical point. Fitted cell probabilities \(\hat p_{ij}\) are:

\begin{verbatim}
R> p11 <- sol$u
R> p21 <- sol$u * sol$r2
R> p12 <- sol$u * sol$s2
R> p22 <- sol$u * sol$r2 * sol$s2
R> p.hat <- matrix_(c(p11, p21, p12, p22), nrow = 2)
\end{verbatim}

\[
\hat p = \frac{1}{\left(y_{11} + y_{12} + y_{21} + y_{22}\right)^{2}}  \left[\begin{matrix}\left(y_{11} + y_{12}\right) \left(y_{11} + y_{21}\right) & \left(y_{11} + y_{12}\right) \left(y_{12} + y_{22}\right)\\\left(y_{11} + y_{21}\right) \left(y_{21} + y_{22}\right) & \left(y_{12} + y_{22}\right) \left(y_{21} + y_{22}\right)\end{matrix}\right]
\]

To verify that the maximum likelihood estimate has been found, we compute the Hessian matrix
which is negative definite (the Hessian matrix is diagonal so the eigenvalues are the diagonal entries and these are all negative), output omitted:

\begin{verbatim}
R> H <- hessian(logL, list(u, r2, s2)) |> simplify()
\end{verbatim}

\hypertarget{example-an-auto-regression-model}{%
\subsection{Example: An auto regression model}\label{example-an-auto-regression-model}}

\hypertarget{symbolic-computations}{%
\subsubsection{Symbolic computations}\label{symbolic-computations}}

In this section we study the auto regressive model of order \(1\) (an AR(1) model), see
e.g. Shumway and Stoﬀer (2016), p.~75 ff.~for details:
Consider random variables \(x_1, x_2, \dots, x_n\) following a stationary zero mean AR(1) process:

\begin{equation}
  x_i = a x_{i-1} + e_i; \quad i=2, \dots, n,
  \label{eq:ar1}
\end{equation}

where \(e_i \sim N(0, v)\) and all \(e_i\)s are independent. Note that \(v\) denotes the variance.
The marginal distribution of \(x_1\) is also assumed normal, and for the process to be stationary
we must have that the variance \(\mathbf{Var}(x_1) = v / (1-a^2)\).
Hence we can write \(x_1 = \frac 1 {\sqrt{1-a^2}} e_1\).

For simplicity of exposition, we set \(n=4\). All terms \(e_1, \dots, e_4\) are independent and \(N(0, v)\) distributed. Let \(e=(e_1, \dots, e_4)\) and \(x=(x_1, \dots x_4)\). Hence \(e \sim N(0, v I)\). Isolating
error terms in \eqref{eq:ar1} gives

\[
  e= \left[\begin{matrix}e_{1}\\e_{2}\\e_{3}\\e_{4}\end{matrix}\right] = \left[\begin{matrix}\sqrt{1 - a^{2}} & . & . & .\\- a & 1 & . & .\\. & - a & 1 & .\\. & . & - a & 1\end{matrix}\right] \left[\begin{matrix}x_{1}\\x_{2}\\x_{3}\\x_{4}\end{matrix}\right] = L x  .
\]

Since
\(\mathbf{Var}(e)=v I\) we have \(\mathbf{Var}(e)=v I=L \mathbf{Var}(x) L^\top\) so the covariance matrix of \(x\) is \(V=\mathbf{Var}(x) = v L^- (L^-)^\top\) while the concentration matrix (the inverse covariance
matrix) is \(K=v^{-1}L^\top L\):

\begin{verbatim}
R> def_sym(a, v)
R> n <- 4
R> L <- diff_mat(n, "-a")
R> L[1, 1] <- sqrt(1-a^2)
R> Linv <- inv(L)
R> K <- crossprod_(L) / v
R> V <- tcrossprod_(Linv) * v
\end{verbatim}

\begin{align} 
    L^{-1} &= \left[\begin{matrix}\frac{1}{\sqrt{1 - a^{2}}} & . & . & .\\\frac{a}{\sqrt{1 - a^{2}}} & 1 & . & .\\\frac{a^{2}}{\sqrt{1 - a^{2}}} & a & 1 & .\\\frac{a^{3}}{\sqrt{1 - a^{2}}} & a^{2} & a & 1\end{matrix}\right] , \\ 
    K &= \frac{1}{v}  \left[\begin{matrix}1 & - a & . & .\\- a & a^{2} + 1 & - a & .\\. & - a & a^{2} + 1 & - a\\. & . & - a & 1\end{matrix}\right] , \\ 
    V &= v  \left[\begin{matrix}\frac{1}{1 - a^{2}} & \frac{a}{1 - a^{2}} & \frac{a^{2}}{1 - a^{2}} & \frac{a^{3}}{1 - a^{2}}\\\frac{a}{1 - a^{2}} & \frac{a^{2}}{1 - a^{2}} + 1 & \frac{a^{3}}{1 - a^{2}} + a & \frac{a^{4}}{1 - a^{2}} + a^{2}\\\frac{a^{2}}{1 - a^{2}} & \frac{a^{3}}{1 - a^{2}} + a & \frac{a^{4}}{1 - a^{2}} + a^{2} + 1 & \frac{a^{5}}{1 - a^{2}} + a^{3} + a\\\frac{a^{3}}{1 - a^{2}} & \frac{a^{4}}{1 - a^{2}} + a^{2} & \frac{a^{5}}{1 - a^{2}} + a^{3} + a & \frac{a^{6}}{1 - a^{2}} + a^{4} + a^{2} + 1\end{matrix}\right]  .
  \end{align}

The zeros in the concentration matrix \(K\) implies a conditional
independence restriction: If the \(ij\)th element of a concentration
matrix is zero then \(x_i\) and \(x_j\) are conditionally independent
given all other variables, see e.g. Højsgaard, Edwards, and Lauritzen (2012), p.~84 for
details.

Next, we take the step from symbolic computations to numerical
evaluations. The joint distribution of \(x\) is multivariate normal
distribution, \(x\sim N(0, K^{-1})\). Let \(W=x x^\top\) denote the
matrix of (cross) products. The log-likelihood is therefore (ignoring
additive constants)
\[ 
\log L = \frac n 2 (\log \mathbf{det}(K) - x^\top K x) = \frac n 2 (\log \mathbf{det}(K) - \mathbf{tr}(K W)), 
\]
where we note that \(\mathbf{tr}(KW)\) is the
sum of the elementwise products of \(K\) and \(W\) since both matrices are
symmetric. Ignoring the constant \(\frac n 2\),
this can be written symbolically to obtain the expression in
this particular case:

\begin{verbatim}
R> x <- vector_sym(n, "x")
R> logL <- log(det(K)) - sum(K * (x %*% t(x))) |> simplify()
\end{verbatim}

\[
\log L = \log{\left(- \frac{a^{2}}{v^{4}} + \frac{1}{v^{4}} \right)} - \frac{- 2 a x_{1} x_{2} - 2 a x_{2} x_{3} - 2 a x_{3} x_{4} + x_{1}^{2} + x_{2}^{2} \left(a^{2} + 1\right) + x_{3}^{2} \left(a^{2} + 1\right) + x_{4}^{2}}{v} .
\]

\hypertarget{numerical-evaluation}{%
\subsubsection{Numerical evaluation}\label{numerical-evaluation}}

Next we illustrate how bridge the gap from symbolic computations to numerical computations based on a dataset:
For a specific data vector we get:

\begin{verbatim}
R> xt <- c(0.1, -0.9, 0.4, 0.0)
R> logL. <- subs(logL, x, xt) 
\end{verbatim}

\[
\log L = \log{\left(- \frac{a^{2}}{v^{4}} + \frac{1}{v^{4}} \right)} - \frac{0.97 a^{2} + 0.9 a + 0.98}{v} .
\]

We can use R for numerical maximization of the likelihood and constraints on the
parameter values can be imposed e.g.~in the \texttt{optim()} function:

\begin{verbatim}
R> logL_wrap <- as_func(logL., vec_arg = TRUE)
R> eps <- 0.01
R> par <- optim(c(a=0, v=1), logL_wrap, 
+              lower=c(-(1-eps), eps), upper=c((1-eps), 10),
+              method="L-BFGS-B", control=list(fnscale=-1))$par
R> par
\end{verbatim}

\begin{verbatim}
#>      a      v 
#> -0.376  0.195
\end{verbatim}

The same model can be fitted e.g.~using R's \texttt{arima()} function as follows (output omitted):

\begin{verbatim}
R> arima(xt, order = c(1, 0, 0), include.mean = FALSE, method = "ML")
\end{verbatim}

It is less trivial to do the optimization in \texttt{caracas} by solving the score equations.
There are some possibilities for putting assumptions on variables
in \texttt{caracas} (see the ``Reference'' vignette), but
it is not possible to restrict the parameter \(a\) to only take values in \((-1, 1)\).

\hypertarget{example-variance-of-average-of-correlated-variables}{%
\subsection{Example: Variance of average of correlated variables}\label{example-variance-of-average-of-correlated-variables}}

Consider random
variables \(x_1,\dots, x_n\) where \(\mathbf{Var}(x_i)=v\) and \(\mathbf{Cov}(x_i, x_j)=v r\) for \(i\not = j\), where \(0 \le |r| \le1\).
For \(n=3\), the covariance matrix of \((x_1,\dots, x_n)\) is therefore

\begin{equation}
  \label{eq:1}
  V = v R = v \left[\begin{matrix}1 & r & r\\r & 1 & r\\r & r & 1\end{matrix}\right]. 
\end{equation}

Let \(\bar x = \sum_i x_i / n\) denote the average. Suppose interest is
in the variance of the average, \(\mathbf{Var}(\bar x)\), when \(n\) goes to
infinity. One approach is as follow: Let \(1\) denote an \(n\)-vector of
\(1\)'s and let \(V\) be an \(n \times n\) matrix with \(v\) on the diagonal
and \(v r\) outside the diagonal. Then \(\mathbf{Var}(\bar x)=\frac 1 {n^2} 1^\top V 1\). The answer lies in studying the limiting behaviour of
this expression when \(n \rightarrow \infty\).
First, we must calculate variance of a sum \(x. = \sum_i x_i\)
which is \(\mathbf{Var}(x.) = \sum_i \mathbf{Var}(x_i) + 2 \sum_{ij:i<j} \mathbf{Cov}(x_i, x_j)\) (i.e., the sum of the elements of the covariance matrix).
We can do this in \texttt{caracas} as follows:

\begin{verbatim}
R> def_sym(v, r, n, j, i) 
R> var_sum <- v * (n + 2 * sum_(sum_(r, j, i + 1, n), i, 1, n - 1)) |> simplify()
R> var_avg <- var_sum / n^2
\end{verbatim}

\[
\mathbf{Var}(x.) = n v \left(r \left(n - 1\right) + 1\right),
\quad
\mathbf{Var}(\bar x) = \frac{v \left(r \left(n - 1\right) + 1\right)}{n}.
\]

From hereof, we can study the limiting behavior of the variance
\(\mathbf{Var}(\bar x)\) in different situations:

\begin{verbatim}
R> l_1 <- lim(var_avg, n, Inf)         ## when sample size n goes to infinity
R> l_2 <- lim(var_avg, r, 0, dir='+')  ## when correlation r goes to zero
R> l_3 <- lim(var_avg, r, 1, dir='-')  ## when correlation r goes to one
\end{verbatim}

Moreover, for a given correlation \(r\) it is instructive to investigate
how many independent variables, say \(k_n\) the \(n\) correlated variables
correspond to (in the sense of the same variance of the average),
because then \(k_n\) can be seen as a measure of the amount of information
in data. Moreover, one might study how \(k_n\) behaves as function of \(n\)
when \(n \rightarrow \infty\). That is we must (1) solve \(v (1 + (n-1)r)/n = v/k\) for \(k\) and (2) find the limit \(l_k = \lim_{n\rightarrow\infty} k_n\):

\begin{verbatim}
R> def_sym(k_n)
R> sol <- solve_sys(var_avg - v / k_n, k_n)
R> k_n <- sol[[1]]$k_n
R> l_k <- lim(k_n, n, Inf)
\end{verbatim}

The findings above are:
\[
l_1 = r v, \quad
l_2 = \frac{v}{n}, \quad
l_3 = v, \quad
k_n = \frac{n}{n r - r + 1}, \quad 
l_k = \frac{1}{r} .
\]

With respect to \(k_n\), it is illustrative to supplement the symbolic
computations above with numerical evaluations, which shows that
even a moderate correlation reduces the effective sample size substantially:

\begin{verbatim}
R> dat <- expand.grid(r=c(.1, .2, .5), n=c(10, 50))
R> k_fn <- as_func(k_n)
R> dat$k_n <- k_fn(r=dat$r, n=dat$n)
R> dat$l_k <- 1 / dat$r
R> dat
\end{verbatim}

\begin{verbatim}
#>     r  n  k_n l_k
#> 1 0.1 10 5.26  10
#> 2 0.2 10 3.57   5
#> 3 0.5 10 1.82   2
#> 4 0.1 50 8.47  10
#> 5 0.2 50 4.63   5
#> 6 0.5 50 1.96   2
\end{verbatim}

\hypertarget{possible-topics-and-projects-for-students}{%
\section{Possible topics and projects for students}\label{possible-topics-and-projects-for-students}}

\begin{enumerate}
\def\labelenumi{\arabic{enumi}.}
\tightlist
\item
  Related to Section {[}Linear models{]}:

  \begin{enumerate}
  \def\labelenumii{\alph{enumii})}
  \tightlist
  \item
    The orthogonal projection
    matrix onto the span of the model matrix \(X\) is \(P=X (X^\top X)^{-1}X^\top\). The residuals are \(r=(I-P)y\). From this one may
    verify that these are not all independent.
  \item
    If one of the factors
    is ignored, then the model becomes a one-way analysis of variance
    model, at it is illustrative to redo the computations in Section
    {[}Linear models{]} in this setting.
  \item
    Likewise if an interaction between the two factors
    is included in the model. What are the residuals in this case?
  \end{enumerate}
\item
  Related to Section {[}Logistic regression{]}:

  \begin{enumerate}
  \def\labelenumii{\alph{enumii})}
  \tightlist
  \item
    In \protect\hyperlink{each-component-of-the-likelihood}{Each component of the
    likelihood}, Newton-Rapson can be implemented to solve the likelihood
    equations and compared to the output from \texttt{glm()}.
    Note how sensitive Newton-Rapson is to starting point.
    This can be solved by another optimisation scheme, e.g.~
    Nelder-Mead (optimising the log likelihood) or BFGS
    (finding extreme for the score function).
  \item
    The example is done as logistic regression with the logit
    link function. Try other link functions such as cloglog (complementary log-log).
  \end{enumerate}
\item
  Related to Section {[}Maximum likelihood under constraints{]}:

  \begin{enumerate}
  \def\labelenumii{\alph{enumii})}
  \tightlist
  \item
    Identifiability of the parameters was handled by not including
    \(r_1\) and \(s_1\) in the specification of \(p_{ij}\). An alternative is
    to impose the restrictions \(r_1=1\) and \(s_1=1\), and this can also
    be handled via Lagrange multipliers. Another alternative is to regard
    the model as a log-linear model where \(\log p_{ij} = \log u + \log r_i + \log s_j = \tilde{u} + \tilde{r}_i + \tilde{s}_j\). This model
    is similar in its structure to the two-way ANOVA for Section {[}Linear
    models{]}. This model can be fitted as a generalized linear model
    with a Poisson likelihood and \(\log\) as link function. Hence, one
    may modify the results in Section {[}Logistic regression{]} to
    provide an alternative way of fitting the model.
  \item
    A simpler task is
    to consider a multinomial distribution with four categories,
    counts \(y_i\) and cell probabilities \(p_i\), \(i=1,2,3,4\) where \(\sum_i p_i=1\). For this model, find the maximum likelihood estimate for
    \(p_i\) (use the Hessian to verify that the critical point is a maximum).
  \end{enumerate}
\item
  Related to Section {[}An \(AR(1)\) model{]}:

  \begin{enumerate}
  \def\labelenumii{\alph{enumii})}
  \tightlist
  \item
    Compare the estimated parameter values with those obtained from
    the \texttt{arima()} function.
  \item
    Modify the model in Equation \eqref{eq:ar1} by
    setting \(x_1 = a x_n + e_1\) (``wrapping around'') and see what happens
    to the pattern of zeros in the concentration matrix.
  \item
    Extend the
    \(AR(1)\) model to an \(AR(2)\) model (``wrapping around'') and
    investigate this model along the same lines. Specifically,
    where are the conditional independencies (try at least \(n=6\))?
  \end{enumerate}
\item
  Related to Section {[}Variance of the average of correlated data{]}: It
  is illustrative to study such behaviours for other covariance
  functions.
  Replicate the calculations for the covariance matrix of the form
  \begin{equation}
    \label{eq:ex5}
    V = v R = v \left[\begin{matrix}1 & r & 0\\r & 1 & r\\0 & r & 1\end{matrix}\right],
  \end{equation}
  i.e., a special case of a Toeplitz matrix.
  How many independent variables, \(k\), do
  the \(n\) correlated variables correspond to?
\end{enumerate}

\hypertarget{discussion-and-future-work}{%
\section{Discussion and future work}\label{discussion-and-future-work}}

We have presented the \texttt{caracas} package and argued that the
package extends the functionality of R significantly with respect to
symbolic mathematics. One practical virtue of \texttt{caracas} is
that the package integrates nicely with \texttt{Rmarkdown},
Allaire et al. (2021), (e.g.~with the \texttt{tex()} functionality) and thus
supports creating of scientific documents and teaching material. As
for the usability in practice we await feedback from users.

Another related package we mentioned in the introduction is \texttt{Ryacas}.
This package has existed for many years and is still of relevance.
\texttt{Ryacas} probably has fewer features than \texttt{caracas}. On the other
hand, \texttt{Ryacas} does not require Python (it is compiled), is faster for
some computations (like matrix inversion). Finally, the Yacas language
(A. Z. Pinkus and Winitzki 2002; A. Pinkus, Winnitzky, and Mazur 2016) is extendable (see e.g.~the vignette
``User-defined yacas rules'' in the \texttt{Ryacas} package).

One possible future development could be an R package which is
designed without a view towards the underlying engine (SymPy or Yacas)
and which then draws more freely from SymPy and Yacas.
In this connection we mention that there are additional resources
on CRAN such as \CRANpkg{calculus} (Guidotti 2022).

Lastly, with respect to freely available resources in a CAS context, we would
like to draw attention to \texttt{WolframAlpha}, see
\url{https://www.wolframalpha.com/}, which provides an online service for
answering (mathematical) queries.

\hypertarget{acknowledgements}{%
\section{Acknowledgements}\label{acknowledgements}}

We would like to thank the R Consortium for financial support for
creating the \texttt{caracas} package, users for pin pointing aspects
that can be improved in \texttt{caracas} and Ege Rubak (Aalborg
University, Denmark), Malte Bødkergaard Nielsen (Aalborg
University, Denmark), Poul Svante Eriksen (Aalborg
University, Denmark), and reviewers
for constructive comments.

\hypertarget{references}{%
\section*{References}\label{references}}
\addcontentsline{toc}{section}{References}

\hypertarget{refs}{}
\begin{CSLReferences}
\leavevmode\vadjust pre{\hypertarget{ref-rmarkdown}{}}%
Allaire, JJ, Yihui Xie, Jonathan McPherson, Javier Luraschi, Kevin Ushey, Aron Atkins, Hadley Wickham, Joe Cheng, Winston Chang, and Richard Iannone. 2021. \emph{Rmarkdown: Dynamic Documents for r}. \url{https://github.com/rstudio/rmarkdown}.

\leavevmode\vadjust pre{\hypertarget{ref-ryacas}{}}%
Andersen, Mikkel Meyer, and Søren Højsgaard. 2019. {``{Ryacas: A computer algebra system in R}.''} \emph{Journal of Open Source Software} 4 (42). \url{https://doi.org/10.21105/joss.01763}.

\leavevmode\vadjust pre{\hypertarget{ref-caracas:21}{}}%
---------. 2021. {``{caracas: Computer algebra in R}.''} \emph{Journal of Open Source Software} 6 (63): 3438. \url{https://doi.org/10.21105/joss.03438}.

\leavevmode\vadjust pre{\hypertarget{ref-JSSv104i05}{}}%
Guidotti, Emanuele. 2022. {``{calculus: High-Dimensional Numerical and Symbolic Calculus in R}.''} \emph{Journal of Statistical Software} 104 (1): 1--37. \url{https://doi.org/10.18637/jss.v104.i05}.

\leavevmode\vadjust pre{\hypertarget{ref-hojsgaard}{}}%
Højsgaard, Søren, David Edwards, and Steffen Lauritzen. 2012. \emph{Graphical Models with {R}}. New York: Springer. \url{https://doi.org/10.1007/978-1-4614-2299-0}.

\leavevmode\vadjust pre{\hypertarget{ref-doBy}{}}%
Højsgaard, Søren, and Ulrich Halekoh. 2023. \emph{{doBy: Groupwise Statistics, LSmeans, Linear Estimates, Utilities}}. \url{https://github.com/hojsgaard/doby}.

\leavevmode\vadjust pre{\hypertarget{ref-mccullagh}{}}%
McCullagh, P, and John A Nelder. 1989. \emph{{Generalized Linear Models}}. 2nd ed. Chapman \& Hall/CRC Monographs on Statistics and Applied Probability. Philadelphia, PA: Chapman \& Hall/CRC.

\leavevmode\vadjust pre{\hypertarget{ref-sympy}{}}%
Meurer, Aaron, Christopher P. Smith, Mateusz Paprocki, Ondřej Čertík, Sergey B. Kirpichev, Matthew Rocklin, AMiT Kumar, et al. 2017. {``SymPy: Symbolic Computing in Python.''} \emph{PeerJ Computer Science} 3 (January): e103. \url{https://doi.org/10.7717/peerj-cs.103}.

\leavevmode\vadjust pre{\hypertarget{ref-nocedal}{}}%
Nocedal, Jorge, and Stephen J. Wright. 2006. \emph{{[}Numerical Optimization}. Springer New York. \url{https://doi.org/10.1007/978-0-387-40065-5}.

\leavevmode\vadjust pre{\hypertarget{ref-Pinkus2002}{}}%
Pinkus, Ayal Z., and Serge Winitzki. 2002. {``{YACAS: A Do-It-Yourself Symbolic Algebra Environment}.''} In \emph{Proceedings of the Joint International Conferences on Artificial Intelligence, Automated Reasoning, and Symbolic Computation}, 332--36. AISC '02/Calculemus '02. London, UK, UK: Springer-Verlag. \url{https://doi.org/10.1007/3-540-45470-5_29}.

\leavevmode\vadjust pre{\hypertarget{ref-yacas}{}}%
Pinkus, Ayal, Serge Winnitzky, and Grzegorz Mazur. 2016. {``Yacas - yet Another Computer Algebra System.''} \url{https://yacas.readthedocs.io/en/latest/}.

\leavevmode\vadjust pre{\hypertarget{ref-shumway:etal:16}{}}%
Shumway, Robert H., and David S. Stoﬀer. 2016. \emph{Time Series Analysis and Its Applications}. Fourth Edition. Springer.

\leavevmode\vadjust pre{\hypertarget{ref-UN17}{}}%
United Nations General Assembly. 2015. {``Sustainable Development Goals.''}

\leavevmode\vadjust pre{\hypertarget{ref-reticulate}{}}%
Ushey, Kevin, JJ Allaire, and Yuan Tang. 2020. \emph{Reticulate: Interface to 'Python'}. \url{https://CRAN.R-project.org/package=reticulate}.

\leavevmode\vadjust pre{\hypertarget{ref-ggplot2}{}}%
Wickham, Hadley. 2016. \emph{Ggplot2: Elegant Graphics for Data Analysis}. Springer-Verlag New York. \url{https://ggplot2.tidyverse.org}.

\leavevmode\vadjust pre{\hypertarget{ref-RMarkdownDefinitiveGuide}{}}%
Xie, Yihui, J. J. Allaire, and Garrett Grolemund. 2018. \emph{R Markdown: The Definitive Guide}. Boca Raton, Florida: Chapman; Hall/CRC. \url{https://bookdown.org/yihui/rmarkdown}.

\leavevmode\vadjust pre{\hypertarget{ref-RMarkdownCookbook}{}}%
Xie, Yihui, Christophe Dervieux, and Emily Riederer. 2020. \emph{R Markdown Cookbook}. Boca Raton, Florida: Chapman; Hall/CRC. \url{https://bookdown.org/yihui/rmarkdown-cookbook}.

\end{CSLReferences}

\appendix

\hypertarget{appendix}{%
\section{Appendix}\label{appendix}}

\hypertarget{installation}{%
\subsection{Installation}\label{installation}}

The \texttt{caracas} package is available on CRAN and can be installed as usual with \texttt{install.packages(\textquotesingle{}caracas\textquotesingle{})}.
Please ensure that you have SymPy installed, or else install it:

\begin{verbatim}
if (!caracas::has_sympy()) {
  caracas::install_sympy()
}
\end{verbatim}

The \texttt{caracas} package uses the \texttt{reticulate} package (to run Python code).
Thus, if you wish to configure your Python environment, you need to
first load \texttt{reticulate}, then configure the Python environment, and at
last load \texttt{caracas}.
The Python environment can be configured as
in \texttt{reticulate}'s ``Python Version Configuration'' vignette.
Again, configuring the Python environment needs to be done before
loading \texttt{caracas}.
Please find further details in \texttt{reticulate}'s documentation.

\hypertarget{low-level-access-to-engines}{%
\subsection{Low-level access to engines}\label{low-level-access-to-engines}}

Since \texttt{caracas} provides essentially an R interface to SymPy using
the \texttt{reticulate} package (Ushey, Allaire, and Tang 2020), everything that can be done
with \texttt{caracas} can also be done directly in Python. In this
connection, it is recommended to refer to SymPy's elaborate
documentation at \url{https://docs.sympy.org}. We illustrate calling Sympy
directly in connection with finding the roots of a polynomial
derivative as done in the \protect\hyperlink{calculus}{calculus} section (the power operator in
Python is \texttt{**} and not \texttt{\^{}} as in R):

\begin{verbatim}
R> library(reticulate)
R> s <- import("sympy") 
R> py_run_string("from sympy import *")
R> py_run_string("x = symbols('x')")
R> p <- py_eval("1 - x**2 + x**3 + x**4/4 - 3 * x**5 / 5 + x**6 / 6")
R> p$evalf(subs = list(x = 1))
\end{verbatim}

\begin{verbatim}
#> 0.816666666666667
\end{verbatim}

\begin{verbatim}
R> sol <- s$solve(s$diff(p, "x"), "x")
\end{verbatim}

We can obtain \texttt{p} in LaTeX format with the command

\begin{verbatim}
R> s$latex(p)
\end{verbatim}

The same can be achieved using standard R syntax with \texttt{caracas} as:

\begin{verbatim}
R> def_sym(x)
R> p <- 1 - x^2 + x^3 + x^4/4 - 3 * x^5 / 5 + x^6 / 6
R> sol <- solve_sys(der(p, x), x)
\end{verbatim}

Another example is from linear algebra: For a matrix \(X\) find \((X^\top X)^{-1}\):

\begin{verbatim}
R> library(reticulate)
R> s <- import("sympy") 
R> py_run_string("from sympy import *")
R> py_run_string("a = symbols('a')")
R> X_str <- "Matrix([[a, 1],[a, 1],[1, 0]])"
R> X <- py_eval(X_str, convert = FALSE) 
R> (X$T * X)$inv()
\end{verbatim}

\begin{verbatim}
#> Matrix([
#> [ 1,         -a],
#> [-a, a**2 + 1/2]])
\end{verbatim}

The same can be achieved using standard R syntax with \texttt{caracas} as:

\begin{verbatim}
R> X <- matrix_(c("a", "1", "a", "1", "1", "0"), nrow = 3, byrow=TRUE)
R> (t(X) %*% X) |> inv() 
\end{verbatim}

As seen, using SymPy directly can be powerful, but it also involves
using text strings and knowing more about Python and SymPy internals
and these requirements steepens the learning curve for both writing
and reading the code.

\hypertarget{extending-caracas}{%
\subsection{\texorpdfstring{Extending \texttt{caracas}}{Extending caracas}}\label{extending-caracas}}

It is possible to easily extend \texttt{caracas} with additional
functionality from SymPy which we illustrate below. This example
illustrates how to use SymPy's \texttt{diff()} function to perform univariate
differentiations multiple times. The partial derivative of \(\sin(xy)\)
with respect to \(x\) and \(y\) is found with \texttt{diff} in SymPy:

\begin{verbatim}
R> sympy <- get_sympy()
R> sympy$diff("sin(x * y)", "x", "y")
\end{verbatim}

\begin{verbatim}
#> -x*y*sin(x*y) + cos(x*y)
\end{verbatim}

One the other hand, the \texttt{der()} function in \texttt{caracas} finds the gradient:

\begin{verbatim}
R> def_sym(x, y)
R> f <- sin(x * y) 
R> der(f, list(x, y))
\end{verbatim}

\begin{verbatim}
#> [c]: [y*cos(x*y)  x*cos(x*y)]
\end{verbatim}

This is a design-choice in \texttt{caracas}. If we want to obtain
the functionality from SymPy
we can write a new function that invokes \texttt{diff} in SymPy using the
\texttt{sympy\_func()} function in \texttt{caracas}:

\begin{verbatim}
R> der_diff <- function(expr, ...){
+    sympy_func(expr, "diff", ...)
+ }
R> der_diff(sin(x * y), x, y)
\end{verbatim}

\begin{verbatim}
#> [c]: -x*y*sin(x*y) + cos(x*y)
\end{verbatim}

This latter function is especially useful if we need to find the higher-order
derivative with respect to the same variable:

\begin{verbatim}
R> sympy$diff("sin(x * y)", "x", 100L)
R> der_diff(sin(x * y), x, 100L)
\end{verbatim}